\documentclass[aps,prl,twocolumn,superscriptaddress]{revtex4}
\usepackage{tikz}
\usepackage[english]{babel}
\usepackage[utf8]{inputenc}
\usepackage{indentfirst}
\usepackage{amsmath}
\usepackage{amssymb}
\usepackage{eufrak}
\usepackage{graphicx}
\usepackage{psfrag}

\newcommand{\lam}{{\boldsymbol \lambda}_M}
\newcommand{\upa}{\uparrow}
\newcommand{\doa}{\downarrow}
\newcommand{\lan}{{\boldsymbol \lambda}_N}
\newcommand{\xil}{{\boldsymbol \xi}_L}

\newcommand{\complex}{\mathbb{C}}
\newcommand{\fa}{\mathfrak{a}}

\newcommand{\eps}{\varepsilon}

\newcommand{\ordo}{\mathcal{O}}
\newcommand{\veff}{v_{\text{eff}}}

\usepackage{amsthm}

\newcommand{\ket}[1]{{\left|#1\right\rangle}}
\newcommand{\bra}[1]{{\left\langle #1\right|}}

\newcommand{\skalarszorzat}[2]{{\langle #1 | #2 \rangle}}

\newcommand{\mRR}{\Theta}

\newcommand{\secc}[1]{\textit{#1}.---}

\usepackage{ifpdf}

\ifpdf
\usepackage{epstopdf}
\usepackage[pdftex,colorlinks,urlcolor=blue,citecolor=blue,linkcolor=blue]{hyperref}
\else
\usepackage[hypertex,colorlinks,urlcolor=blue,citecolor=blue,linkcolor=blue]{hyperref}
\fi
\begin{document}

\title{
  Algebraic construction of current operators in integrable spin chains
}
\author{Bal\'azs Pozsgay}
\affiliation{MTA-BME Quantum Dynamics and Correlations Research Group,\\
    Department of Theoretical Physics,\\ Budapest University
of Technology and Economics,\\ 1521 Budapest, Hungary}

\begin{abstract}
Generalized Hydrodynamics is a recent theory that describes the large scale transport properties of
one dimensional integrable models. At the heart of this theory lies an exact quantum-classical
correspondence, which states that the flows
of the conserved quantities are essentially quasi-classical even in the interacting quantum many
body models. We provide the algebraic background to this observation, by embedding the current
operators of the integrable spin chains into the canonical framework of Yang-Baxter integrability.
Our construction can be applied in a large variety of models including the XXZ spin chains, the Hubbard model,
and even in models lacking particle conservation such as the XYZ chain. 
Regarding the XXZ chain we present a simplified proof of the recent exact results for the current
mean values, and explain how their quasi-classical nature emerges from the exact computations.
\end{abstract}

\maketitle

\secc{Introduction}
The non-equilibrium dynamics of one dimensional  quantum integrable systems has attracted a lot of
interest \cite{nonequilibrium-intro-review}.
Integrable models possess a large number of
commuting conserved charges, constraining their dynamical processes and leading to dissipationless
and factorized scattering.
This exotic dynamical behaviour has a number of experimentally measurable consequences, for example
a lack
of thermalization \cite{QNewtonCradle,gge-experiment1}.
Two central theoretical problems have been the equilibration in isolated integrable models, and the
description of transport  in spatially
inhomogeneous and/or driven systems. Regarding equilibration it is now
accepted that the emerging steady states can be described by the Generalized Gibbs Ensemble
\cite{JS-CGGE,rigol-quench-review}. Regarding
transport the theory of Generalized Hydrodynamics
(GHD) was introduced in \cite{doyon-ghd,jacopo-ghd}, which describes both the ballistic modes and
also the diffusive corrections
\cite{doyon-jacopo-ghd-diffusive,jacopo-benjamin-bernard--ghd-diffusion-long}.  Recent works
\cite{vasseur-superdiff,vir-superdiff,enej-superdiff} also treated the phenomenon of super-diffusion.

In GHD a central role is played by the current operators describing the flow of conserved
quantities. The continuity relations for these flows completely
determine the transport at the Euler-scale \cite{doyon-ghd,jacopo-ghd}. It is thus of utmost
importance to understand the mean currents in local or global equilibria. The works
\cite{doyon-ghd,jacopo-ghd} argued that in the thermodynamic limit the currents are  given by
a formula of the type
\begin{equation}
  \label{Jeredeti}
  J=\int d\lambda\ \rho(\lambda) \veff(\lambda) h(\lambda),
\end{equation}
where $\lambda$ is a rapidity parameter, $h(\lambda)$ is the
one-particle charge eigenvalue, $\rho(\lambda)$ is the differential particle density per volume
and rapidity, and $\veff(\lambda)$ is an ``effective velocity'' that describes the propagation of
single particle wave packets in the presence of the other particles \cite{flea-gas}. Clearly, this
concept is quasi-classical, and it assumes the dissipationless scattering of integrable models.

The formula \eqref{Jeredeti} has received continued attention. It was known that it holds in models
equivalent to free bosons or free fermions, where
$\veff(\lambda)=e'(\lambda)/p'(\lambda)$ is the group velocity
\cite{maurizio-currents}.
In interacting cases proofs were given in various settings 
\cite{doyon-ghd,kluemper-spin-current,dinh-long-takato-ghd,bajnok-vona-currents,spohn-toda-proof,takato-spohn}.
The paper \cite{sajat-currents} derived a new and exact finite volume formula for the mean currents
in the Heisenberg spin chains,
and a connection to long range
deformed models was pointed out in
\cite{sajat-longcurrents}.
However, the microscopic proofs were not transparent enough and did not fully explain
why there exist such simple and exact formulas for the 
currents. Furthermore,
the direct algebraic representation of the current operators was missing. 

In this Letter we fill this gap. We make a direct connection to the Quantum Inverse Scattering
Method (QISM) pioneered by L. Faddeev and the Leningrad school
\cite{Korepin-Book,faddeev-history}. This is the canonical framework to treat  local quantum integrable
systems. For the first time we show that the QISM also accommodates the current operators, leading to a simplified
rigorous derivation of their mean values, corroborating their quasi-classical nature.

\secc{Charges and currents}
We consider integrable spin chains in finite volume, given by a local Hamiltonian $\hat H$ acting on the Hilbert space
 $\mathcal{H}=\otimes_{j=1}^L V_j$ with $V_j\simeq \complex^d$. We assume periodic boundary
 conditions.

Examples are the XXX, XXZ and XYZ
 Heisenberg spin chains \cite{Korepin-Book,Baxter-Book}, or the 1D Hubbard 
 model \cite{Hubbard-Book}.
These integrable models possess a canonical set of local
conserved charges $\hat Q_\alpha$ that are in involution $[\hat Q_\alpha,\hat Q_\beta]=0$, such that
$\hat H$ belongs to the family. The charges can be written as $\hat Q_\alpha=\sum_x \hat
q_\alpha(x)$, with $\hat q_\alpha(x)$  being the charge density 
operators. 

The flow of these charges is described by the current operators $\hat J_\alpha(x)$, defined through the
continuity relations
\begin{equation}
  \label{Ja}
i  \left[\hat H, \hat q_\alpha(x)\right]=\hat J_{\alpha}(x)-\hat J_{\alpha}(x+1).
\end{equation}
Following \cite{benjamin-takato-note-ghd,sajat-currents} we also introduce the generalized current
operators $\hat J_{\alpha,\beta}$ that describe the flow of $\hat Q_\alpha$ under the time evolution
generated by $\hat Q_\beta$. They are defined through 
\begin{equation}
  \label{Jab}
i  \left[\hat Q_\beta, \hat q_\alpha(x)\right]=\hat J_{\alpha,\beta}(x)-\hat J_{\alpha,\beta}(x+1).
\end{equation}
It is our goal to compute the exact mean values of $\hat J_{\alpha,\beta}$ in the eigenstates of the
models, and to show that they always take a form analogous to \eqref{Jeredeti}.

\secc{Transfer matrices}
The standard method to find the commuting set of charges is the QISM
\cite{Korepin-Book,faddeev-history}. 
We start with the so-called $R$-matrix $R(\mu,\nu)\in
End(\complex^d\otimes \complex^d)$ which satisfies the Yang-Baxter relation:
\begin{equation}
  \label{YB}
  \begin{split}
     R_{12}(\lambda_{1},\lambda_2)&R_{13}(\lambda_1,\lambda_3)R_{23}(\lambda_2,\lambda_3)=\\
&  =R_{23}(\lambda_2,\lambda_3) R_{13}(\lambda_1,\lambda_3) R_{12}(\lambda_{1},\lambda_2).
  \end{split}
\end{equation}
This is a relation for operators acting on the triple tensor product $V_1\otimes V_2\otimes V_3$
and we assume $V_j\simeq \complex^d$. It is understood that each $R_{jk}$ acts only on the
corresponding vector spaces. Examples for $R$-matrices (describing the above mentioned models)
can be found in
\cite{Baxter-Book,Korepin-Book,Hubbard-Book}.
We assume that the so-called regularity and unitarity conditions hold:
\begin{equation}
  \label{inv}
  \begin{split}
    R(\lambda,\lambda)&=P\\
  R_{12}(\lambda_1,\lambda_2)R_{21}(\lambda_2,\lambda_1)&=1.
  \end{split}
 \end{equation}
Here  $P$ is the permutation operator and $R_{21}(u,v)=PR_{12}(u,v)P$.

The charges are obtained from a commuting set of transfer matrices.
Let us take an auxiliary space $V_a\simeq \complex^d$ and the Lax-operators
$\mathcal{L}_{a,j}(u)$ which act on $V_a$ and on a local space $V_j$ with $j=1\dots L$, where $L$ is
the length of the chain. We require that the following exchange relation holds:
\begin{equation}
  \label{RLL}
  \begin{split}
  R_{b,a}(\nu,\mu) & \mathcal{L}_{b,j}(\nu)  \mathcal{L}_{a,j}(\mu)= \mathcal{L}_{a,j}(\mu) \mathcal{L}_{b,j}(\nu)    R_{b,a}(\nu,\mu)
\end{split}
\end{equation}
with $a,b$ referring to two different auxiliary spaces. 
It follows from \eqref{YB} that
  $\mathcal{L}_{a,j}(\mu)=R_{a,j}(\mu,\xi_0)$
is a solution to \eqref{RLL}, where $\xi_0$ is a fixed parameter of the model. In the following we
use this choice and assume that $\xi_0=0$.

The monodromy matrix acting on $V_a\otimes \mathcal{H}$ is defined as
\begin{equation}
  \label{ttdef}
  \hat T_a(\mu)=\mathcal{L}_{a,L}(\mu)\dots\mathcal{L}_{a,1}(\mu).
\end{equation}
The transfer matrix is its partial trace over the auxiliary space:
$\hat  t(\mu)=\text{Tr}_a \hat  T_a(\mu)$.
The fundamental exchange relations \eqref{RLL} guarantee that
  $[\hat t(\mu),\hat t(\nu)]=0$.
A generating function for global charges is then defined as \cite{Korepin-Book,faddeev-history}
\begin{equation}
  \label{Qnudef}
 \hat  Q(\nu)\equiv (-i)\hat t^{-1}(\nu)\frac{d}{d\nu} \hat t(\nu)
\end{equation}
The traditional charges are the Taylor coefficients:
\begin{equation}
\hat   Q(\nu)=\sum_{\alpha=2}^\infty \frac{\nu^{\alpha-2}}{(\alpha-2)!}\hat Q_\alpha.
\end{equation}
The $\hat Q_\alpha$ are extensive, and the density
$\hat q_\alpha(x)$ spans $\alpha$ sites \cite{Luscher-conserved}; in particular $\hat H\sim \hat Q_2$. The definition
\eqref{Qnudef} makes sense in any finite volume, but it gives 
the correct $\hat Q_\alpha$ only if $L>\alpha$. In the $L\to\infty$ limit the operator $\hat Q(\mu)$ is expected to be
quasi-local in some neighborhood of $\mu=0$, for proofs in concrete cases see
\cite{prosen-xxx-quasi,prosen-enej-quasi-local-review,sajat-su3-gge}. 

\secc{Charge densities}
Writing $\hat Q(\mu)=\textstyle \sum_{x=1}^L  \hat q(\mu,x)$ we can identify
the corresponding operator density as
\begin{equation}
  \label{qmuxdef}
  \begin{split}
 \hat  q(\mu,x)&\equiv (-i)\hat t^{-1}(\mu)\times\\
  &  \times \text{Tr}_a    \left[ \hat T_a^{[L,x+1]}(\mu)    
      \partial_\mu {\mathcal{L}}_{a,x}(\mu)    \hat T_a^{[x-1,1]}(\mu) \right].
  \end{split}    
\end{equation}
Here we defined the partial monodromy matrices acting on a segment $[x_1\dots x_2]$
as
\begin{equation}
\hat   T^{[x_2,x_1]}_a(\mu)=\mathcal{L}_{a,x_2}(\mu) \dots \mathcal{L}_{a,x_1}(\mu).
\end{equation}
The definition \eqref{qmuxdef} is homogeneous in space:
 $\hat  q(\mu,x)=\hat U^{-1}\hat q(\mu,x+1)\hat U$,
where $\hat U$ is the cyclic shift operator to the right.

\secc{Current operators}
We also construct a generating function for the currents:
\begin{equation}
  \label{Jsum}
\hat   J(\mu,\nu,x)=\sum_{\alpha=2}^\infty \sum_{\beta=2}^\infty\frac{\mu^{\alpha-2}}{(\alpha-2)!}
  \frac{\nu^{\beta-2}}{(\beta-2)!}  \hat J_{\alpha,\beta}(x).
\end{equation}
This two-parameter family of operators satisfies the generalized
continuity relation
\begin{equation}
  \label{Jmunudef}
  i  \left[\hat Q(\nu),\hat q(\mu,x)\right]=  \hat J(\mu,\nu,x)-\hat J(\mu,\nu,x+1).
\end{equation}
The summation in \eqref{Jsum} only makes sense in the $L\to\infty$ limit, where we expect that
$J(\mu,\nu,x)$ is a finite norm operator localized around $x$, at least in some neighborhood of
$\mu=\nu=0$. Relation \eqref{Jmunudef} is well defined in any finite volume, if we use
\eqref{Qnudef}-\eqref{qmuxdef}. 

It is our goal to give an explicit construction for $\hat J(\mu,\nu,x)$. We start with the
commutator 
\begin{equation}
  \label{kolpa}
  \begin{split}
    [\hat t(\nu),&\hat q(\mu,x)]    =(-i)\hat   t^{-1}(\mu)\times\\
   & \times\frac{d}{d\eps} \text{Tr}_{ab}
  \left(\hat T_b(\nu)\hat T^\eps_a(\mu)-\hat T^\eps_a(\mu)\hat T_b(\nu)\right),
  \end{split}
\end{equation}
where now $a$ and $b$ refer to two different auxiliary spaces, and $\hat T^\eps_a(\mu)$ is a
deformed monodromy matrix defined as 
\begin{equation}
  \hat T_a^\eps(\mu)=\hat T_a^{[L,x+1]}(\mu)           {\mathcal{L}}_{a,x}(\mu+\eps)    \hat T_a^{[x-1,1]}(\mu).
\end{equation}
The modification of the rapidity parameter at site $x$ is the reason for the non-commutativity, and
this will result in the appearance of the current operators. 

At $\eps=0$ the intertwining of the monodromy matrices is performed by a repeated application of
\eqref{RLL}.
In $\hat T^\eps(\mu)$ the difference is that there is one Lax operator with a modified rapidity. At that particular site
the exchange is also given by \eqref{RLL}, but it involves $R_{b,a}(\nu,\mu+\eps)$.
Inserting these commutation relations into \eqref{kolpa} and performing the $\eps$-derivative we
eventually obtain
\begin{equation}
  \label{opop3c}
\hat t^{-1}(\nu)  \left[\hat t(\nu),\hat q(\mu,x)\right]=
 \hat \Omega(\mu,\nu,x)-\hat \Omega(\mu,\nu,x-1),
\end{equation}
where we introduced a new ``double row'' operator
\begin{equation}
  \label{Omegadef}
  \begin{split}
  \hat \Omega(\mu,\nu,&x)=
\hat t^{-1}(\nu)\hat t^{-1}(\mu) \text{Tr}_{ab}\left[ \hat T_a^{[L,x+1]}(\mu)\right.\times \\
&\left.\times   \hat T_b^{[L,x+1]}(\nu) \mRR_{a,b}(\mu,\nu)  \hat  T_a^{[x,1]}(\mu)  \hat T_b^{[x,1]}(\nu)\right].
  \end{split}
\end{equation}
Here
\begin{equation}
  \mRR_{a,b}(\mu,\nu)= (-i)  R_{b,a}(\nu,\mu) \partial_\mu   R_{a,b}(\mu,\nu)
\end{equation}
is an operator insertion acting only on the auxiliary spaces,
coupling the two monodromy matrices.

Taking a further $\nu$-derivative on the l.h.s. of \eqref{opop3c} we recognize the continuity equation
\eqref{Jmunudef} and identify
\begin{equation}
  \label{JmunuabOmega}
  \hat J(\mu,\nu,x)=-\hat t(\nu) \partial_\nu\hat \Omega(\mu,\nu,x-1) \hat t^{-1}(\nu).
\end{equation}
Let $\ket{\Psi}$ be an arbitrary eigenstate of the commuting transfer matrices.
For the mean values we get:
\begin{equation}
  \label{JOmega}
   \bra{\Psi}\hat J(\mu,\nu,x)\ket{\Psi}=-\partial_\nu    \bra{\Psi}\hat \Omega(\mu,\nu,x-1)\ket{\Psi}.
 \end{equation}
This connects the $\nu$-derivatives of $\hat\Omega(\mu,\nu,x)$ to the current mean values. To complete
the picture, we also 
compute the initial value at $\nu=0$. Direct substitution
and the regularity condition lead to
$\hat \Omega(\mu,0,x)=\hat q(\mu,x)$.
Thus $\hat \Omega$ not only describes all (generalized) currents, but also all charge
densities. Together with \eqref{JOmega} this is 
the first central result of our work. 

\secc{Symmetry} We discuss the symmetry of $\hat \Omega(\mu,\nu,x)$ under the exchange of
its rapidity variables. The partial monodromy matrices in the definition \eqref{Omegadef} can be
exchanged using \eqref{RLL}. Direct computation shows that $\hat \Omega(\mu,\nu,x)=\hat \Omega(\nu,\mu,x)$ iff
\begin{equation}
   \partial_\mu R_{b,a}(\nu,\mu) +\partial_\nu   R_{b,a}(\nu,\mu)=0.
\end{equation}
This is satisfied if the $R$-matrix is of difference form:
$R_{b,a}(\nu,\mu)=R_{b,a}(\nu-\mu)$. Examples are
the various Heisenberg spin chains, and a famous counter-example is the Hubbard model.
This exchange symmetry results in equalities between different charge and current
operators, as already observed in \cite{sajat-currents}.

\secc{Inhomogeneous cases}
The nature of the operator $\hat \Omega$ is better understood if we also consider
the inhomogeneous spin chains. Let us take generic complex numbers $\xil$ and define the
inhomogeneous monodromy matrix
\begin{equation}
  \label{ttdefinhom}
  \hat T_a(\mu)=R_{a,L}(\mu,\xi_L)\dots R_{a,1}(\mu,\xi_1),
\end{equation}
In this case we can still define the $\hat \Omega$ operator with formula \eqref{Omegadef},
replacing each local Lax operator with their inhomogeneous 
versions, and keeping the insertion $\mRR_{a,b}(\mu,\nu)$ the same.

Even though $\hat \Omega$ is quite complicated, 
there is a remarkable simplification when the parameters $\mu,\nu$ are chosen from the
set $\xil$.
Let us take for simplicity $\mu=\xi_1$, $\nu=\xi_2$ and
set $x=2$. A straightforward computations leads to
\begin{equation}
  \label{xi1xi2}
  \hat \Omega(\xi_1,\xi_2,2)=\mRR_{1,2}(\xi_1,\xi_2).
\end{equation}
This means that for these special values $\hat \Omega(\mu,\nu,x)$ becomes an ultra-local operator acting only
on the first two sites. 
This bridges a connection to the theory of factorized correlation functions
in the XXZ chain
\cite{boos-korepin-first-factorization,hgs2,HGSIII,XXZ-finite-T-factorization,kluemper-goehmann-finiteT-review,kluemper-discrete-functional},
where the mean value of $\mRR_{1,2}(\xi_2,\xi_1)$ is one of the basic building blocks. Our
contribution here is the construction of $\hat \Omega(\mu,\nu,x)$ for general $\mu,\nu$, and
the explanation that it describes the currents and the charges.
The result \eqref{xi1xi2} is also analogous to the 
``solution of the inverse problem'' 
\cite{goehmann-korepin-inverse,maillet-terras-inverse}, where the monodromy matrix elements can be
specialized such that they become ultra-local operators acting on single sites only.

\secc{Mean values}
We return to the homogeneous case and employ a trick originally developed in
\cite{XXZ-finite-T-factorization}. We relate the mean values of $\hat \Omega(\mu,\nu,x)$ to a
transfer matrix eigenvalue in an 
auxiliary problem.
Consider an enlarged spin chain with two extra sites. Choose a
rapidity $\mu$ and a deformation parameter $\eps$. 
The enlarged monodromy
matrix acts on $V_a\otimes V_{L+2}\otimes V_{L+1}\otimes \mathcal{H}$ and is given by
\begin{equation}
  \label{Tpluszdef}
  \hat T^+_a(u)=
R_{a,L+2}(u,\mu+\eps)  R^{t_{L+1}}_{L+1,a}(\mu,u)
  T_a(u),
\end{equation}
where $T_a(u)$ is given by \eqref{ttdef}, and $t_{L+1}$ denotes partial
transposition with respect to the physical space at site $x=L+1$.
The Yang-Baxter
relation implies that $R^{t_{L+1}}_{L+1,a}(\mu,u)$ also satisfies the exchange relation \eqref{RLL},
thus the transfer matrices defined as $\hat t^+(u)=\text{Tr}_a T^+_a(u)$ form a commuting set. 

At $\eps=0$ the extra two sites become decoupled: If $\ket{\Psi}$ is an eigenstate of the
original $\hat t(u)$ with eigenvalue $\Lambda(u)$, then 
\begin{equation}
  \label{decoup}
  \hat t^+(u)\Big(  \ket{\delta}\otimes \ket{\Psi}\Big)=
\Lambda(u)  \Big(  \ket{\delta}\otimes \ket{\Psi}\Big).
\end{equation}
Here $\ket{\delta}$ is the ``delta-state'' given by components $\delta_{ij}$ in the computational
basis. 

After switching on a non-zero $\eps$  the first two sites will affect
the eigenvalues and the eigenvectors. Let $\Lambda^+(u|\mu,\eps)$ be the eigenvalue of $\hat t^+(u)$ on
a state $\ket{\Psi^+}$ 
which in the limit $\eps\to 0$ becomes $\ket{\delta}\otimes \ket{\Psi}$.
A standard first order perturbation theory computation gives
\footnote{Supplemental  Materials to ``Algebraic construction of current operators in integrable spin chains''}
\begin{equation}
  \label{OmegaLambda}
  \bra{\Psi}\hat\Omega(\mu,\nu,x)\ket{\Psi}=
 i \left. \frac{d}{d\eps} \log \Lambda^+(\nu|\mu,\eps)\right|_{\eps=0}.
\end{equation}
This is the second central result of our work, which applies essentially to ``all'' Yang-Baxter
integrable local chains. The eigenvalues $\Lambda^+(\nu|\mu,\eps)$ can always
be found by standard methods of integrability, and this explains why there exist simple exact formulas for
the current mean values. 
The specifics of the model come into play 
only when we are actually solving the auxiliary problem.

\secc{Heisenberg spin chain}
As an example we take the easy-axis XXZ chain defined by the Hamiltonian density
\begin{equation}
  \hat h(j)=  \hat \sigma_j^x  \hat \sigma_{j+1}^x+  \hat \sigma_j^y  \hat \sigma_{j+1}^y
  +\Delta   (\hat \sigma_j^z  \hat \sigma_{j+1}^z-1)
\end{equation}
Here $\hat \sigma^{x,y,z}_j$ are Pauli matrices acting on site $j$ and $\Delta=\cosh(\eta)>1$ is the
anisotropy parameter. The associated $R$-matrix is of the form
\begin{equation}
  \label{RXXZ}
  R(\mu,\nu)=
  \begin{pmatrix}
    1 & 0 & 0 & 0\\
    0 & b(\mu-\nu) & c(\mu-\nu) & 0 \\
    0 & c(\mu-\nu) & b(\mu-\nu) & 0 \\
    0 & 0 & 0 & 1
  \end{pmatrix}.
\end{equation}
with $b(u)=\sin(u)/\sin(u+i\eta)$, $c(u)=\sin(i\eta)/\sin(u+i\eta)$.
  
The model can be solved by the Algebraic Bethe Ansatz (ABA) \cite{Korepin-Book}.
Eigenstates  are labeled by a set of rapidities $\lan$, describing $N$ interacting spin
waves, satisfying the Bethe
equations
\begin{equation}
  \label{Betheorig}
  p(\lambda_k)L+\sum_{j\ne k}^N\delta(\lambda_k-\lambda_j)=2\pi Z_k,\quad Z_k\in \mathbb{Z},
\end{equation}
where $L$ is the length of the chain, and
\begin{equation}
    e^{i p(\lambda)}= \frac{\sin(\lambda-i\eta/2)}{\sin(\lambda+i\eta/2)},\quad
e^{i\delta(\lambda)}=\frac{\sin(\lambda+i\eta)}{\sin(\lambda-i\eta)}.
\end{equation}
For the generating function of the conserved charges
we find the eigenvalues
$\hat Q(\nu)\ket{\lan}  
=Q(\nu) \ket{\lan}$
where
$Q(\nu)\simeq \sum_{j=1}^N h(\lambda_j-\nu)$ and $h(u)=p'(u)$. Here and in the following the $\simeq$ sign means
that there are correction terms behaving as $\ordo(\nu^L)$ or $\ordo(\mu^L)$ for small $\mu,\nu$.

The auxiliary spin chain problem defined by \eqref{Tpluszdef}
can also be solved using ABA. Here we present the outline of the computation; for the details we refer to
\cite{Note1}.
It turns out that the main effect of the extra two sites is that they act as a
momentum dependent twist operator for the particles of the original chain. This
deforms the Bethe equations and their solutions. We get
\begin{equation}
\label{momtwist}
  - \eps h(\lambda_k-\mu)+
  p(\lambda_k)L+\sum_{j\ne k}^N\delta(\lambda_k-\lambda_j)\simeq 2\pi Z_k,
\end{equation}
where $\mu$ is the external parameter introduced in \eqref{Tpluszdef}.
Furthermore, we have $\partial_\nu \log \Lambda^+(\nu|\mu,\eps)\simeq i Q(\nu)$, where $Q(\nu)$ is
the same function introduced above, but 
evaluated at the $\eps$-deformed rapidities \cite{Note1}.
Equations \eqref{JmunuabOmega} and \eqref{OmegaLambda} then lead to
\begin{equation}
  \label{Jxx}
  \bra{\lan}\hat J(\mu,\nu,x)\ket{\lan}\simeq \sum_{j=1}^N h'(\lambda_j-\nu)  \frac{d\lambda_j}{d\eps}.
\end{equation}
If we regard the solution $\lan$ of \eqref{Betheorig} as functions of the $Z_k$, then 
the $\eps$-derivatives can be expressed as
 \begin{equation}
   \frac{d\lambda_j}{d\eps}\simeq \sum_{k=1}^N \frac{\partial\lambda_j}{\partial (2\pi Z_k)}
   h(\lambda_k-\mu).
 \end{equation}
Then the result \eqref{Jxx} is written as
\begin{equation}
  \begin{split}
    \bra{\lan}\hat J(\mu,\nu,x)\ket{\lan}
   &\simeq \sum_{k=1}^N \frac{\partial Q(\nu)}{\partial (2\pi Z_k)}   h(\lambda_k-\mu).
  \end{split}
 \end{equation} 
Expanding to low orders in $\mu$ and $\nu$ we get the final result
 \begin{equation}
   \label{mostgeneral}
  \begin{split}
    \bra{\lan}\hat J_{\alpha,\beta}(x)\ket{\lan}
   &=\sum_{k=1}^N \frac{\partial Q_\beta}{\partial (2\pi Z_k)}   h_\alpha(\lambda_k).
  \end{split}
\end{equation} 
Even though the intermediate formulas were only approximate, the final result
\eqref{mostgeneral} is exact, and agrees with \cite{sajat-currents,sajat-longcurrents}; the exact
formula for $\bra{\lan}\hat\Omega(\mu,\nu,x)\ket{\lan}$ 
is presented in \cite{Note1}.

\secc{Interpretation}
Consider the semi-classical picture of $N$ particles moving on the circle of circumference
$L$, subject to two-particle scattering events described by the phase shift $\delta(\lambda)$ defined
above.
In this situation
$(2\pi Z_k)/L$ can be interpreted as the ``dressed momentum'' of the particles,
which takes into account the interaction between the particles.
Then the formula \eqref{mostgeneral} is interpreted as
\begin{equation}
   \bra{\lan}\hat J_{\alpha,\beta}(x)\ket{\lan}
   =\frac{1}{L}\sum_{k=1}^N v_{\text{eff},\beta}(\lambda_k)   h_\alpha(\lambda_k)
\end{equation}
with $v_{\text{eff},\beta}(\lambda_k)=L\partial Q_\beta/\partial (2\pi Z_k)$ being the
natural generalization of the group velocity under time evolution dictated by $\hat Q_\beta$. For
more details see \cite{sajat-currents,sajat-longcurrents}.

\secc{Thermodynamic limit}
It is possible to take the thermodynamic limit of 
\eqref{mostgeneral} with a direct approach,
 reproducing the results of  \cite{doyon-ghd,jacopo-ghd}.
 Alternatively, we can apply the Quantum
Transfer Matrix approach  \cite{XXZ-finite-T-factorization,kluemper-discrete-functional} directly in
the thermodynamic limit.
 These computations will
be presented elsewhere. 

\secc{Discussion}
We constructed a generating function for the charge densities and the current operators using
standard tools of Yang-Baxter integrability. The main formulas are model independent.

Our construction explains why there exist simple formulas for the current mean values: because they
are tied to certain transfer matrix eigenvalues through \eqref{JOmega} and \eqref{OmegaLambda}.
In integrable models such eigenvalues are always ``easy'' to compute, in contrast with generic
correlation functions, which are much more difficult to handle. This means that the current
operators are the ``next simplest'' operators after the charge densities.

We demonstrated on the example of the XXZ chain that the current mean values have a quasi-classical
interpretation. 
Our derivations suggest that this is a generic feature of integrable spin chains. 
The ultimate physical reason for this behaviour is the dissipationless and
factorized scattering in integrable models, and our work provided new algebraic tools to treat this
phenomenon.
We stress that our computations are completely rigorous. The approximations above were
only introduced to
provide a more intuitive understanding. Thus we made an important step towards proving the emergence
of hydrodynamics in a quantum many body situation.

In future work we plan to compute the currents in models not yet considered in the literature. A
particularly interesting case is the XYZ model, which belongs to the class of models treated here,
but lacks particle conservation on the microscopic level.
Furthermore, it would be interesting to consider
current operators also in the Separation of Variables approach
\cite{sklyanin-sov-review,maillet-sov-uj-1,gromov-separation-mean,niccoli-terras-sov}.

\secc{Acknowledgments}
The author is grateful to Frank G\"ohmann, Benjamin Doyon, Yunfeng Jiang, M\'arton Kor\-mos and
G\'abor Tak\'acs  for useful discussions and  important suggestions about
the manuscript. Furthermore we thank Levente Pristy\'ak for checking of some of the
formulas in the Supplemental Materials.
This research was supported by the BME-Nanotechnology FIKP grant (BME FIKP-NAT),
by the National Research Development and Innovation Office (NKFIH) (K-2016 grant no. 119204), by the
J\'anos Bolyai Research Scholarship of the Hungarian 
Academy of Sciences, and
by the \'UNKP-19-4 New National Excellence Program of the Ministry for Innovation and Technology.

%


\widetext
\newpage

\begin{center}
  \textbf{\large Supplemental Materials:}
  \medskip

  \textbf{\large Algebraic construction of current operators in integrable spin chains}
\end{center}
\setcounter{equation}{0}
\setcounter{figure}{0}
\setcounter{table}{0}
\setcounter{page}{1}
\makeatletter
\renewcommand{\theequation}{S\arabic{equation}}
\renewcommand{\thefigure}{S\arabic{figure}}
\renewcommand{\bibnumfmt}[1]{[S#1]}
\renewcommand{\citenumfont}[1]{S#1}

\section{Relating $\hat \Omega(\mu,\nu)$ to transfer matrix eigenvalues}

Here we prove relation (26) of the main text, which connects the mean values of $\hat
\Omega(\mu,\nu)$ to transfer matrix eigenvalues of an auxiliary spin chain. The derivations in this
Sections are model independent, and they are a simple generalization of an analogous computation
presented in \cite{XXZ-finite-T-factorizationS,kluemper-discrete-functionalS}.

We consider the transfer matrix $\hat t^+(u|\mu,\eps)=\text{Tr}_a \hat T_a^+(u)$, where $\hat T_a^+(u)$
is the monodromy matrix of the auxiliary chain defined in (24) of the main text. It is easy to see
that at $\eps=0$  a subset of eigenstates of $\hat t^+(u|\mu,\eps)$ is given by states of the form
\begin{equation}
\ket{\Psi^+}= \frac{1}{\sqrt{d}} \ket{\delta}\otimes \ket{\Psi},
\end{equation}
where $\ket{\delta}$ is the delta-state on the extra two sites and $\ket{\Psi}$ is an eigenstate of
the original model. The number $d$ is the dimensionality of the local Hilbert space, thus the
two-site vector $\ket{\delta}/\sqrt{d}$ is normalized to 1.

Let us now compute the following action on such a vector:
\begin{equation}
  \label{e1}
  \begin{split}
 \frac{-i}{\sqrt{d}\Lambda(\mu)\Lambda(\nu)}  \hat t^+(\mu|\mu,0)  
\left.\frac{d  \hat t^+(\nu|\mu,\eps)}{d\eps}\right|_{\eps=0} 
\left(\ket{\delta}\otimes \ket{\Psi} \right).
\end{split}
\end{equation}
Here $\Lambda(u)$ with $u=\mu,\nu$ are the eigenvalues of the original transfer matrices on
$\ket{\Psi}$, but at $\eps=0$ they coincide with the eigenvalues of $\hat t(u)$ on $\ket{\Psi^+}$.

Substituting the special form of the transfer matrices, and using the regularity conditions it can
be seen that the resulting vector is equal to
\begin{equation}
  \label{e11}
-\frac{1}{\sqrt{d}}\ket{\delta}\otimes  \Big(\hat\Omega(\mu,\nu)\ket{\Psi}\Big).
\end{equation}
Here an extra minus sign appeared because the transfer matrices in \eqref{e1} lead to the formula (17) of
the main text with $\Theta_{a,b}(\mu,\nu)$ replaced by
\begin{equation}
  (-i)\partial_\mu R_{b,a}(\nu,\mu) R_{a,b}(\mu,\nu)=  i
  R_{b,a}(\nu,\mu) \partial_\mu R_{a,b}(\mu,\nu)=-\Theta_{a,b}(\mu,\nu).
\end{equation}
The first equality above follows from the unitarity condition.

Now we intend to apply the Hellmann-Feynman theorem. Let $\bra{\Psi^+}$ be the dual vector to
$\ket{\Psi^+}$ defined above, normalized to $\skalarszorzat{\Psi^+}{\Psi^+}=1$.  We take the scalar
product with the vector in \eqref{e1}: 
\begin{equation}
  \begin{split}
    &    \frac{-i}{\Lambda(\mu)\Lambda(\nu)}   \bra{\Psi^+} \hat t^+(\mu|\mu,0)
\left.\frac{d  \hat t^+(\nu|\mu,\eps)}{d\eps}\right|_{\eps=0} 
    \ket{\Psi^+}=   \frac{-i}{\Lambda(\nu)}   \bra{\Psi^+}  \left.\frac{d  \hat
        t^+(\nu|\mu,\eps)}{d\eps}\right|_{\eps=0}  \ket{\Psi^+}.
  \end{split}  
\end{equation}
Applying the Hellmann-Feynman theorem 
\begin{equation}
  \label{e2}
    \bra{\Psi^+}\left.\frac{d  \hat t^+(\nu|\mu,\eps)}{d\eps}\right|_{\eps=0}  \ket{\Psi^+}=
\left.\frac{d}{d\eps} \Lambda^+(\nu|\mu,\eps)\right|_{\eps=0}.
 \end{equation}
 Putting \eqref{e1}, \eqref{e11} and \eqref{e2} together
 \begin{equation}
\frac{1}{\sqrt{d}}\bra{\Psi^+}\left[ \ket{\delta}\otimes  \Big(\hat\Omega(\mu,\nu)\ket{\Psi}\Big)\right]=
i \left.\frac{d}{d\eps} \log \Lambda^+(\nu|\mu,\eps)\right|_{\eps=0}.
\end{equation}
To complete the proof we need to show that the l.h.s. above is equal to
$\bra{\Psi}\hat\Omega(\mu,\nu)\ket{\Psi}$. Note that $\bra{\Psi^+}\ne
\ket{\Psi^+}^\dagger$, because
the transfer matrix $\hat t^+(u|\mu,0)$ does not involve Hermitian operators at finite
$\mu$. Nevertheless it can be argued that
\begin{equation}
  \label{e3}
  \frac{1}{\sqrt{d}}\bra{\Psi^+}\left[ \ket{\delta}\otimes
    \Big(\hat \Omega(\mu,\nu)\ket{\Psi}\Big)\right]=
  \bra{\Psi}\hat \Omega(\mu,\nu)\ket{\Psi}.
\end{equation}
To see this, consider $ \ket{\delta}\otimes\Big(\hat \Omega(\mu,\nu)\ket{\Psi}\Big)$, which is a
linear combination of vectors of type 
$\ket{\delta}\otimes \ket{n}$, where $\ket{n}$ are eigenstates of the original transfer matrix. All
these vectors are eigenvectors of $\hat t^+(u|\mu,0)$, and thus orthogonal to $\bra{\Psi^+}$ except
when $\ket{n}=\ket{\Psi}$. This means that when we take scalar product
with $\bra{\Psi^+}$ only the contribution of $\ket{\Psi^+}$ remains, proving \eqref{e3}.

Putting everything together we obtain relation (26) of the main text.

\section{The exact mean values of $\hat\Omega(\mu,\nu)$}

Here we consider the example of the finite volume XXZ spin chain with $\Delta>1$, and compute the
mean values of $\hat\Omega(\mu,\nu,x)$ using relation (26) of the main text.
In this model the Lax operators are equal to the
fundamental $R$-matrix. For the sake of completeness we consider here 
the completely inhomogeneous case. The diagonalization is performed using standard steps of
Algebraic Bethe Ansatz \cite{Korepin-BookS}.

We thus consider the auxiliary spin chain of length $L+2$. Let us spell out the $\eps$-deformed monodromy matrix:
\begin{equation}
  \label{monodromyj}
 \hat T^+_a(u)=R_{L+2,a}(u-\mu-\eps) R^{t}_{L+1,a}(\mu-u)
 R_{L,a}(u-\xi_L)\dots R_{1,a}(u-\xi_1).
\end{equation}
Here we used that $R$ is of the difference form, and $\xil$ are the inhomogeneity parameters. The
transfer matrix is $\hat t^+(u|\mu,\eps)=\text{Tr}_a \hat T^+_a(u)$; we are interested in its eigenvalues
$\Lambda^+(u|\mu,\eps)$. It can be shown that the $\Lambda^+(u|\mu,\eps)$ do not depend on the
ordering on the $\xil$ or the position of the insertion of the extra two sites. Thus the mean values
of  $\hat\Omega(\mu,\nu,x)$ are independent of $x$.

As usual, we decompose the monodromy matrix as
\begin{equation}
  \hat T^+(u)=
  \begin{pmatrix}
   \hat A(u) & \hat B(u) \\  \hat C(u) & \hat D(u)
  \end{pmatrix},
\end{equation}
where now $\hat A(u),\hat B(u),\hat C(u),\hat D(u)$ are operators acting on the spin chain.

We choose the reference state 
\begin{equation}
  \ket{0}_{L+2}=\ket{\upa\doa}\otimes \ket{\upa\dots \upa}_{L}.
\end{equation}
It can be seen that this satisfies the standard annihilation property $\hat
C(u)\ket{0}_{L+2}=0$. Furthermore, the reference state is an eigenvector of $\hat A(u)$ and $\hat
D(u)$ with the eigenvalues
\begin{equation}
  a(u)=\frac{\sin(u-\mu)}{\sin(u-\mu-i\eta)} 
  \qquad
  d(u)=\frac{\sin(u-\mu-\eps)}{\sin(u-\mu-\eps+i\eta)}  \prod_{j=1}^L \frac{\sin(u-\xi_j)}{\sin(u-\xi_j+i\eta)},
\end{equation}
which can be read off the explicit construction \eqref{monodromyj}.

Bethe states $\ket{\lam}$ are created as
\begin{equation}
  \ket{\lam}=\prod_{j=1}^M \hat B(\lambda_j-i\eta/2)\ket{0}_{L+2}.
\end{equation}
The shift of $-i\eta/2$ is introduced for later convenience. They are eigenstates of the transfer
matrix if they satisfy the Bethe equations:
\begin{equation}
  \frac{d(\lambda_j)}{a(\lambda_j)}\prod_{k\ne j}
  \frac{\sin(\lambda_j-\lambda_k+i\eta)}{\sin(\lambda_j-\lambda_k-i\eta)}=1,\quad j=1\dots M.
\end{equation}
In these cases the eigenvalue is
\begin{equation}
  \label{La1}
 \Lambda^+(u|\mu,\eps)=a(u)\prod_{j=1}^M \frac{\sin(u-\lambda_j-i\eta/2)}{\sin(u-\lambda_j+i\eta/2)}
  +d(u)\prod_{j=1}^M \frac{\sin(u-\lambda_j+3i\eta/2)}{\sin(u-\lambda_j+i\eta/2)}.
\end{equation}

We are interested in eigenstates of the auxiliary chain which in the $\eps\to 0$ limit become states
of the form
\begin{equation}
  \ket{\delta}\otimes \ket{\lan},
\end{equation}
where $\ket{\delta}$ is the delta-state positioned at the first two sites, and $\ket{\lan}$ is an
eigenstate of the original spin chain. The delta-state has a non-trivial structure as opposed to the
reference state, and it can be seen that at $\eps=0$ it is created by $\hat B(\mu)$. Thus at finite
$\eps$ we are looking for a set of rapidities with $M=N+1$ given by
\begin{equation}
  \label{lambdat}
  \{\tilde \mu+i\eta/2,\tilde\lambda_1,\dots,\tilde\lambda_N\},
\end{equation}
such that in the $\eps\to 0$ limit they approach $\{\mu+i\eta/2\}\cup \lan$. 
We expect the solution in the form 
\begin{equation}
  \tilde\mu=\mu+\eps\gamma +\ordo(\eps^2),\qquad \tilde\lambda_j=\lambda_j+\eps (\Delta\lambda_j)+\ordo(\eps^2),
\end{equation}
where $\gamma$ is an unknown ratio of the shifts. In the following we concentrate only on the linear terms in $\eps$.

In order to find the solution at $\eps\ne 0$ we substitute \eqref{lambdat} into the Bethe
equations. 
The equation for $\tilde\mu$ becomes
\begin{equation}
  \label{beirva}
  \begin{split}
  \frac{\sin(\tilde\mu-\mu-i\eta)}{\sin(\tilde\mu-\mu)}
  \frac {\sin(\tilde\mu-\mu-\eps)}{\sin(\tilde\mu-\mu-\eps+i\eta)}
  \prod_{j=1}^L \frac{\sin(\tilde\mu-\xi_j)}  {\sin(\tilde\mu-\xi_j+i\eta)}
   \prod_{j=1}^N  \frac{\sin(\tilde\mu-\tilde\lambda_j+3i\eta/2)}{\sin(\tilde\mu-\tilde\lambda_j-i\eta/2)}&=1.
  \end{split}
\end{equation}
This equation fixes $\gamma$. Note that both the numerator and the denominator have a factor that
goes to 0 in the $\eps\to 0$ limit, thus $\gamma$ can be found if we substitute $\eps=0$ in the remaining factors.
Let us introduce the function
\begin{equation}
  \fa(u)=
    \prod_{j=1}^L \frac{\sin(u-\xi_j)}{\sin(u-\xi_j+i\eta)}
   \prod_{j=1}^N  \frac{\sin(u-\lambda_j+3i\eta/2)}{\sin(u-\lambda_j-i\eta/2)},
 \end{equation}
 where $\lambda_j$ are solutions to the original Bethe equations, given in the main text.
Then \eqref{beirva} gives 
\begin{equation}
  \gamma=\frac{\fa(\mu)}{1+\fa(\mu)}.
\end{equation}
Going further, the shifts $\Delta \lambda_k$ can be obtained from the logarithmic form of the Bethe
equations for the $\tilde\lambda_k$:
\begin{equation}
  \label{beirva2}
  p(\tilde\lambda_k-\mu-\eps)+p(\tilde\lambda_k-\mu-i\eta)+\delta(\tilde\lambda_k-\tilde\mu-i\eta/2)+
  \sum_{j=1}^L p(\tilde\lambda_k-\xi_j)  +\sum_{j\ne k} \delta(\tilde\lambda_j-\tilde\lambda_k)=2\pi Z_k.
\end{equation}
Here $p(u)$ is the single particle lattice momentum and $\delta(u)$ is the scattering phase shift
defined in the main text.  Using the explicit formulas it can be seen that the sum of the first
three terms on the l.h.s. vanishes if $\eps=0$ and $\tilde\mu=\mu$. 

In order to compute the shift in the set $\lan$ we introduce the Gaudin matrix $G$, given by
\begin{equation}
G_{jk}=\left.\frac{\partial (2\pi Z_k)}{\partial \lambda_j}\right|_{\eps=0}.
\end{equation}
More explicitly
\begin{equation}
  G_{jk}=\delta_{jk}\left[\sum_{l=1}^L h(\lambda_k-\xi_l)+\sum_{l=1}^N \varphi(\lambda_j-\lambda_l)\right]
-\varphi(\lambda_j-\lambda_k),
\end{equation}
where we introduced
\begin{equation}
  h(u)=p'(u)=\frac{\sinh(\eta)}{\sin(u+i\eta/2)\sin(u-i\eta/2)},\qquad
  \varphi(u)=\delta'(u)=-\frac{\sinh(2\eta)}{\sin(u+i\eta)\sin(u-i\eta)}.
\end{equation}
It is important the first three terms in \eqref{beirva2} do not contribute to the Gaudin matrix,
because the sum of their $\lambda$-derivatives is only $\ordo(\eps)$, and we have set $\eps=0$.

Then the shift vector $\Delta {\boldsymbol \lambda}$ is found from
\begin{equation}
  G \cdot \Delta{\boldsymbol\lambda}={\bf H}(\mu),
\end{equation}
where ${\bf H}(\mu)$ is a vector of length $N$ given by the components
\begin{equation}
  H_k(\mu)=h(\tilde\lambda_k-\mu)+\frac{\fa(\mu)}{1+\fa(\mu)}\varphi(\tilde\lambda_k-\mu-i\eta/2).
\end{equation}
This follows from the l.h.s. of \eqref{beirva2} after substituting also $\tilde\mu=\mu+\eps \fa(\mu)/(1+\fa(\mu))$.

Let us spell out the eigenvalue given by \eqref{La1}:
\begin{equation}
  \label{La2}
  \begin{split}
  \Lambda^+(\nu|\mu,\eps)=&\frac{\sin(\nu-\mu)}{\sin(\nu-\tilde\mu)} \frac{\sin(\nu-\tilde\mu-i\eta)}{\sin(\nu-\mu-i\eta)}
  \prod_{j=1}^N \frac{\sin(\nu-\tilde\lambda_j-i\eta/2)}{\sin(\nu-\tilde\lambda_j+i\eta/2)}+\\
  &  +\frac{\sin(\nu-\mu-\eps)}{\sin(\nu-\tilde\mu)}
  \frac{\sin(\nu-\tilde\mu+i\eta)}  {\sin(\nu-\mu-\eps+i\eta)}
  \prod_{j=1}^L
\frac{\sin(\nu-\xi_j)}{\sin(\nu-\xi_j+i\eta)}
  \prod_{j=1}^N \frac{\sin(\nu-\tilde\lambda_j+3i\eta/2)}{\sin(\nu-\tilde\lambda_j+i\eta/2)}.
  \end{split}
\end{equation}
At $\eps=0$, $\tilde\mu=\mu$ and $\tilde\lambda_k=\lambda_k$ this coincides with the usual transfer
matrix eigenvalue function of 
the original chain of length $L$. 

The mean value of $\hat\Omega(\mu,\nu,x)$ is found from the derivative
\begin{equation}
\bra{\lan}\hat \Omega(\mu,\nu,x)\ket{\lan}=i
  \left.  \frac{1}{\Lambda^+(\nu|\mu,\eps)} \frac{d\Lambda^+(\nu|\mu,\eps)}{d\eps}\right|_{\eps=0}.
\end{equation}
A straightforward, but somewhat lengthy computation yields
\begin{equation}
  \label{Omexact}
  \begin{split}
\bra{\lan}\hat \Omega(\mu,\nu,x)\ket{\lan}=
  {\bf H}(\nu) \cdot G^{-1}\cdot {\bf H}(\mu)+
l(\mu,\nu)+l(\nu,\mu)
  \end{split}
\end{equation}
with
\begin{equation}
  l(\mu,\nu)=\frac{h(\nu-\mu+i\eta/2)}{(1+\fa(\mu))(1+\fa^{-1}(\nu))}
\end{equation}
Eq. \eqref{Omexact} is the exact finite volume result.

The mean values of the currents  are obtained after taking
the homogeneous limit $\xi_j=0$ and performing an expansion
into a Taylor series in $\mu,\nu$. In a finite volume we are only interested in the current
operators $\hat J_{\alpha,\beta}$ with
$\alpha+\beta\le L$, because only these operators fit into the volume. 
In the homogeneous case $\fa(u)\sim u^L$,
therefore it is safe to substitute $\fa(u)\approx 0$ in the above formulas. This means that we can
approximate ${\bf  H}(\mu)\approx {\bf h}(\mu)$ where ${\bf h}(\mu)$ is a vector with elements
$h(\lambda_j-\mu)$. Furthermore, we can substitute $l(\mu,\nu)\approx 0$. This leads to the approximate formulas
presented in the main text.

The two main effects of this approximation are that in \eqref{La2} we
neglect the second term (as usually when computing the charge eigenvalues), and in the exact Bethe
equations \eqref{beirva2} we substitute  $\tilde\mu=\mu$, which leads to the approximation
\begin{equation}
p(\tilde\lambda_k-\mu-\eps)-p(\tilde\lambda_k-\mu)+
  \sum_{j=1}^L p(\tilde\lambda_k-\xi_j)  +\sum_{j\ne k}
  \delta(\tilde\lambda_j-\tilde\lambda_k)\simeq 2\pi Z_k.
\end{equation}
This leads to eq. (31) of the main text. In this approximation the mean value of $\hat\Omega$ is
\begin{equation}
  \label{Omegakb}
  \begin{split}
\bra{\lan}\hat \Omega(\mu,\nu,x)\ket{\lan}\simeq
 {\bf h}(\nu) \cdot G^{-1}\cdot {\bf h}(\mu).
  \end{split}
\end{equation}
Taking a $\nu$-derivative and expanding to low orders in $\mu,\nu$ we get the exact result
\begin{equation}
  \label{Jabab}
  \bra{\lan}\hat J_{\alpha,\beta}(x)\ket{\lan}=
  {\bf h}_\beta' \cdot G^{-1}\cdot {\bf h}_\alpha.
\end{equation}
Here  ${\bf h}_\alpha$ is a vector of length $N$ with elements $h_\alpha(\lambda_j)$ where
\begin{equation}
  h_\alpha(\lambda)=\left.\left(\frac{\partial }{\partial \mu}\right)^{\alpha-2} h(\lambda-\mu)\right|_{\mu=0}
\end{equation}
and similarly for ${\bf h}_\beta'$:
\begin{equation}
   h_\beta'(\lambda)=\left.\frac{\partial}{\partial \lambda}\left(\frac{\partial }{\partial \mu}\right)^{\beta-2} h(\lambda-\mu)
\right|_{\mu=0}=-h_{\beta+1}(\lambda).
\end{equation}
Formula \eqref{Jabab} is the result found in \cite{sajat-currentsS,sajat-longcurrentsS}. It is equivalent to eq. (36)
of the main text, because
\begin{equation}
   ({\bf h}_\beta' \cdot G^{-1})_k=
   \sum_{j=1}^N h'_\beta(\lambda_j)\frac{\partial \lambda_j}{\partial (2\pi Z_k)}=
\frac{\partial Q_\beta}{\partial (2\pi Z_k)}.
\end{equation}

\end{document}